\documentclass[10pt]{article}

\usepackage{graphicx}
\usepackage{amsfonts}
\usepackage{amsmath}
\usepackage{amssymb}
\usepackage{mathrsfs} 
\usepackage{lipsum}
\usepackage{color}
\usepackage{siunitx}
\usepackage{authblk}
\usepackage[margin=2cm]{geometry}
\usepackage{mathtools, cuted}

\usepackage{bm}
\usepackage{amsmath}

\usepackage{color}
\definecolor{BDcolor}{rgb}{1,0,0}

\definecolor{Manuecolor}{rgb}{0,0,1}

\begin{document}

\title{Marginal regeneration-induced drainage of surface bubbles}
\author{Jonas Miguet$^1$, Marina Pasquet$^2$, Florence Rouyer $^3$, Yuan Fang $^4$, Emmanuelle Rio$^2$\\
}
\maketitle

\begin{abstract}
The prediction of the lifetime of surface bubbles necessitates a better understanding of the thinning dynamics of the bubble cap. In 1959, Mysel \textit{et al.} \cite{mysels1959soap}, proposed that \textit{marginal regeneration} \textit{i.e.}  the rise of patches, thinner than the film should be taken into account to describe the film drainage.
Nevertheless, an accurate description of these buoyant patches and of their dynamics as well as a quantification of their contribution to the thinning dynamics is still lacking. In this paper, we visualize the patches,  and show that their rising velocities and sizes are in good agreement with models respectively based on the balance of gravitational and surface viscous forces and on a Rayleigh-Taylor like instability \cite{Seiwert2017,Shabalina2019}. 
Our results suggest that, in an  environment saturated in humidity, the drainage induced by their dynamics correctly describes the film drainage at the apex of the bubble within the experimental error bars. We conclude that the film thinning of soap bubbles is indeed controlled, to a large extent, by \textit{marginal regeneration} in the absence of evaporation. 
\end{abstract}

\maketitle

\section{Introduction}

What is the lifetime of a surface soap bubble? 
This question contributes to the fascination that bubbles exert on people whatever their age \cite{Salkin2015, Seychelles2008} and the answer is crucial to understand exchanges between liquid baths and their environment.
This concerns various domains such as sparkling beverages industry for the dispersion of flavors  \cite{Liger-Belair2009}, climate models for the exchange between oceans and atmosphere \cite{Boucher2013,murphy1998} or dispersion models for pollutants above swimming pools \cite{PoulainPRL2018}. 
Despite this wide range of applications, the question of the bubble stability remains largely open. This is directly linked to the global thinning dynamics of the bubble cap, which requires to take into account two contributions: the evaporation and the liquid flow within the film (drainage).

The effect of evaporation has been dismissed in most experimental studies concerning bubbles lifetime and film thinning dynamics and in most of them, the atmospheric humidity is measured \textit{a posteriori} rather than controlled. 
It is only recently that this parameter has been properly  controlled and incorporated in the models \cite{PoulainJFM2018,PoulainPRL2018,Miguet2020}.

The liquid drainage between the interfaces is dictated by a balance between capillary suction from the thin film to the meniscus, gravity for Bo $\geq$ 10 \cite{Miguet2020} and viscous dissipation. The capillary suction is controlled by the shape of the bubble, more specifically by the curvature gradient at the bottom of the film and is controlled at the macroscopic scale by the balance between the buoyancy force and the surface tension. On the other hand, the viscous dissipation is set by the ability of the gas/liquid interfaces to oppose a resistance to the flow of the liquid. In the case of bare bubbles, the flow is extensional and the velocity is therefore limited by the viscous dissipation induced by the velocity gradients along the flow. If the liquid is viscous enough, the process is slow enough and can be observed more easily, leading to exponential decay of the film thickness at the apex \cite{Debregeas1998,kovcarkova2013,howell1999draining}. In the case of low viscosity, the rupture of bare bubbles is extremely fast \cite{PoulainJFM2018}.

In practice, however, aqueous liquids always feature surface active impurities \cite{Lhuissier2012} 
and surface tension gradients emerge upon the penetration of bubbles through the interface and subsequent stretching of the surface. These gradients allow the interfaces to carry the film weight \cite{Gennes2001} and to oppose \textit{some} resistance to the liquid flow, therefore leading to more stable systems. If this resistance is high enough the velocity profile is expected to lead to a Poiseuille-flow profile within the film.
However, in the most common situation of standard surfactants neither the plug-flow nor the Poiseuille-flow profiles allow to describe the experiments \cite{Champougny2016,Bhamla2017}.

\begin{figure}[!ht]
  \centering
   \includegraphics[width=0.7\textwidth]{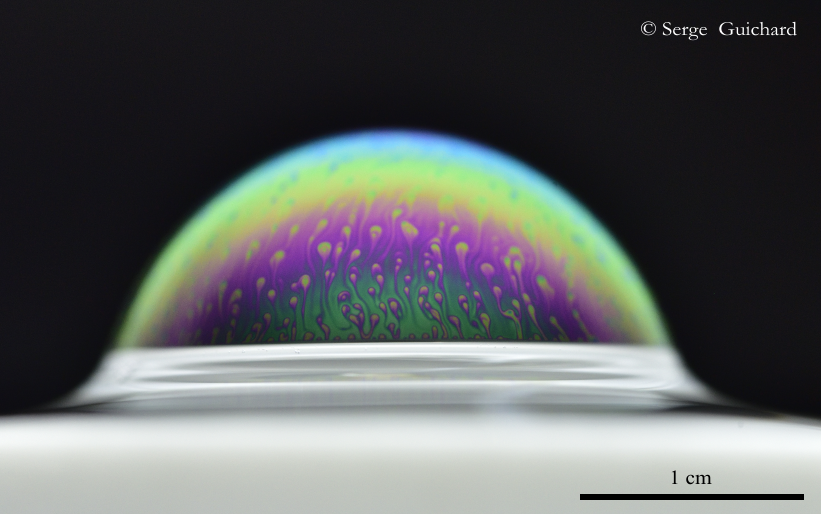}
 \caption{Photograph of a surface bubble with a radius of curvature of 1.13 cm (Bo=25) illuminated in white light for a TTAB concentration of 0.5 cmc, credits: Serge Guichard.}
 \label{fig:Photo_bulle_couleurs}
\end{figure}

In the case of flat films, another scenario, deducted from the observation of the whole film dynamics and of thickness heterogeneity in the presence of surfactants, has been proposed by Mysels \textit{et al.} \cite{mysels1959soap} and relies on the so-called \textit{marginal regeneration}.
This phenomenon is linked to the apparition of thin patches near the borders, in the vicinity of the meniscus (Fig. \ref{fig:Photo_bulle_couleurs}). The drainage mechanism associated assumes that exchanges between thick patches, which enter the meniscus and thin patches, which are extracted out of it, regulate the global thinning of the film.
Gros \textit{et al.} \cite{gros2021marginal} proposed to name this phenomenon \textit{sliding puzzle dynamics} since it necessitates a conservation of the surface area. 

The literature on these striking patches (Fig. \ref{fig:Photo_bulle_couleurs}) remains surprisingly scarce \cite{Stein1991,Bruinsma1995,Aradian2001,Nierstrasz1998, Nierstrasz1999,naire2004,Lhuissier2012,Seiwert2017,gros2021marginal}
To explain their presence, Mysels \textit{et al.} proposed the existence of a marginal pinch or dimple due to a difference of capillary pressure between the film and the meniscus \cite{mysels1959soap}. 
A theoretical model predicting the pinching dynamics in flat films, has been proposed by Aradian \textit{et al} \cite{Aradian2001} but has never been verified experimentally. 

The patches come from the destabilization of the marginal pinch \cite{Bruinsma1995}. 
Two different scenarios are available in the literature, which rely either on a Benard-Marangoni like instability due to the presence of surface tension gradients \cite{Lhuissier2012,Nierstrasz1998,Nierstrasz1999, Nierstrasz2001} or on a Rayleigh-Taylor like instability driven by gravity \cite{Shabalina2019}.
A similar gravity-driven scenario can also be observed in slightly different experiments, in which a thick film is created above a thinner one \cite{goldstein2014instability}.
In both scenarios, the unstable nature of the pinching triggers the extraction of patches that still have a thickness comparable to that of the main film \cite{Lhuissier2012,gros2021marginal}.
This has been confirmed by Nierstrasz and Frens, who performed experimental measurements of the patch thickness on planar vertical films by interferometry and found a value of 0.8 for the ratio between the patches thickness and that of the rest of the film. This is comparable to the upper limit found for this ratio in the model of Gros \textit{et al.} \cite{gros2021marginal} equal to 0.7. a result of the order of the averaged film thickness (the patch thickness is found equal to 0.8 times the film thickness) \cite{Nierstrasz1998}, which is comparable to the upper limit found in the model of Gros \textit{et al.} \cite{gros2021marginal} equal to 0.7.

The   question   of   the quantitative and precise  contribution   of   the marginal   regeneration   to   the   film   drainage  also remains largely  open. Following the proposition by Mysels \textit{et al.} \cite{mysels1959soap}, Seiwert \textit{et al.} \cite{Seiwert2017}  proposed that, in planar vertical films, the rising of thin patches, which replace the thick ones is the main mechanism responsible for drainage. Lhuissier \textit{et al.} \cite{Lhuissier2012} proposed a second scenario to link the bubble cap drainage to marginal regeneration with a drainage limited by the liquid flow through the pinch.
This model leads to a prediction, which has been proven to describe well the experimental data \cite{PoulainPRL2018,Miguet2020}.
Our main result, which we will discuss in the following, is that both scenarios hold, which is probably due to a coupling through the pinch thickness. 

The purpose of this article is to explore in detail the question of the contribution of the sliding puzzle dynamics to bubble drainage. To tackle this problem, we first observe and characterize marginal regeneration patches. We then perform a comparison between direct thinning measurements at the apex of the bubble cap and a model allowing to derive the contribution of the marginal regeneration to drainage.  As mentioned earlier, this necessitates to suppress the influence of evaporation on thinning and thus, to  control the atmospheric humidity. Our conclusion is  that the sliding puzzle dynamics is a sufficient ingredient to explain the film drainage in absence of evaporation.

\begin{figure}[htb!]
 \centering
   \includegraphics[width=1\textwidth]{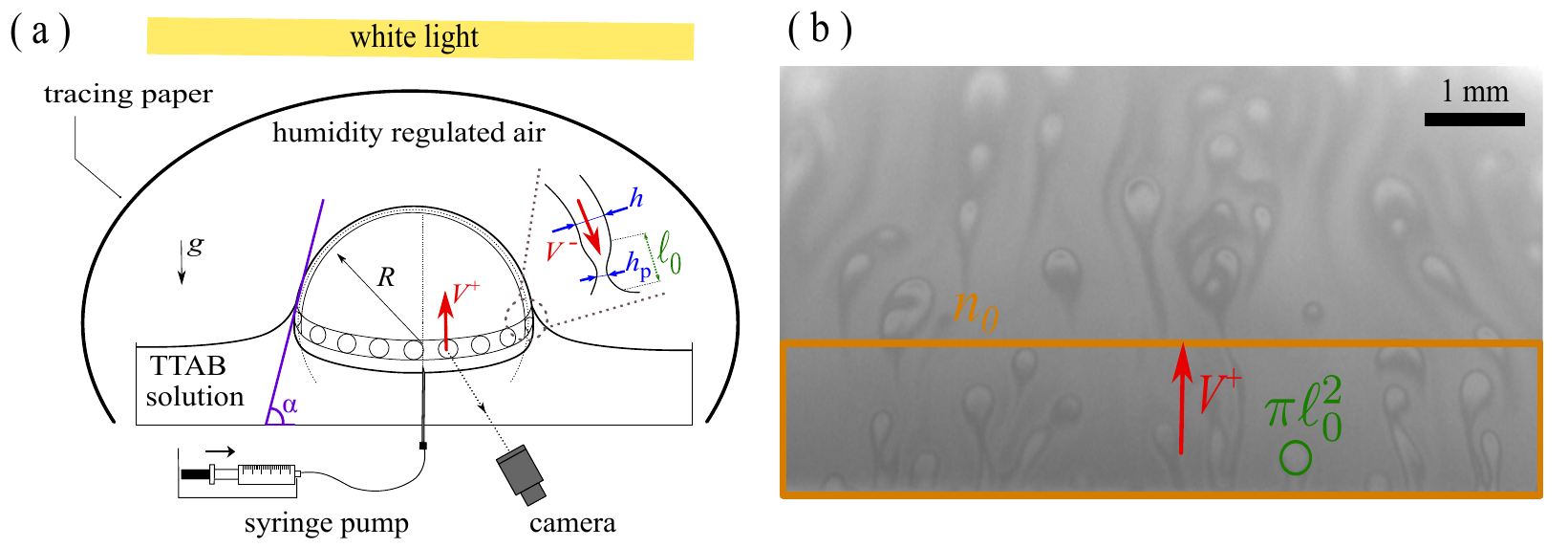}
 \caption{(a) Sketch of the experimental setup with the notations used. (b) Example of an experimental image centered on the bottom of the bubble on which we can measure the characteristic size of the patches $\ell_0$ (m) and the number of patches per unit area $n_0$ (m$^{-2}$). }
 \label{fig:Schéma_dispositif_experimental-Crop_bulle_monochrome}
\end{figure}

\section{Patches visualization and characterization}

The study is conducted on bubbles stabilized by an aqueous surfactant solution made with recrystallized TTAB (tetradecyl trimethylammonium bromide, purchased from Sigma-Aldrich) at a concentration of 0.5 cmc (critical micelle concentration, equal to 3.6 mmol.L$^{-1}$)  and ultrapure water (resistivity = 18.2 M$\Omega$.cm). 
The TTAB recrystallization  \cite{Stubenrauch2005} ensures a good reproducibility of the data.  
The bubbles are generated by injecting air below the surface, through a vertical Teflon bevelled needle (outer diameter = 0.56 mm). The flow rate of the injected air is controlled by a syringe pump and fixed at Q = 15 mL$\cdot$min$^{-1}$ (Fig.\ref{fig:Schéma_dispositif_experimental-Crop_bulle_monochrome} (a)). 
The bubble size is controlled by the injection time and the radius $R$ of the emerged spherical cap is comprised between 9 and 13 mm. The temperature is comprised between 20 and 23$^{\mathcal{O}}$C.
In our study, $\gamma$ = 50 mN$\cdot$m$^{-1}$ and $\rho$ = 1000 kg$\cdot$m$^{-3}$ so that the Bond number Bo=$\rho g R^2/ \gamma$, which controls the shape adopted by the bubble at the interface \cite{Nguyen2013}, is always larger than 5.
The entire setup is enclosed in a box, in which the humidity is controlled with a humidity regulator developed by F. Boulogne \cite{Boulogne2019}.
In these conditions, we observe systematically some marginal regeneration after a few tens of seconds.

To characterize the patches size and dynamics, we perform direct visualization. We illuminate uniformly the bubbles with a white light panel 
above the tank containing the solution and a diffusing ``tunnel'' of tracing paper placed around the bubble (Fig. \ref{fig:Schéma_dispositif_experimental-Crop_bulle_monochrome}(a)). 
A monochromatic camera (Basler acA3800 - 14um) at the entrance of the ``tunnel'' allows to record the images, at a framerate of 10 images per second.
The bubble size and the contact angle $\alpha$ between the bath surface and the bubble cap just above the meniscus is defined in Fig. \ref{fig:Schéma_dispositif_experimental-Crop_bulle_monochrome} and is measured for each bubble on the images. The angle $\alpha$ varies between 55$^{\mathcal{O}}$ and 65$^{\mathcal{O}}$, in good agreement with Teixeira \textit{et al.} \cite{Teixeira2015}.
The characteristic size of the patches $\ell_0$, their number per unit area $n_0$ and their rising velocity $V^{+}$ are extracted from the images using the ImageJ software  (Fig.\ref{fig:Schéma_dispositif_experimental-Crop_bulle_monochrome} (b)).
These measurements are performed in the vicinity of the meniscus to avoid distortions due to the inclination of the surface. The contrast is unfortunately too poor and too variable for an automatic detection, thus the patches' size is measured by superimposing a circle on the image and by extracting its radius, $n_0$ is counted by hand in a box of approximately 1700$\times$200 pixels$^2$ and $V^{+}$ is estimated by averaging the displacement around a given time for at least 5 successive frames.
To carry out these measurements, averages were taken to ensure the robustness of the values: for a given time, the size of the patches, their rising velocity and their average number per unit area were measured 10 times.
We also measured the film thickness $h$ at the apex of the bubbles using a white light spectrometer \cite{Champougny2016,Miguet2020}. 
The spectra obtained are semi-automatically processed making use of a Python code using the library Oospectro \cite{Oospectro2019}.

\begin{figure}[htb]
 \centering
   \includegraphics{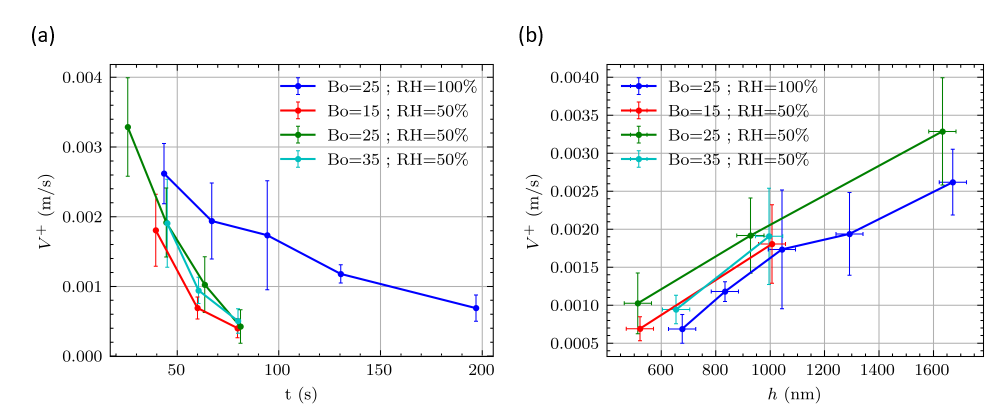}
 \caption{(a) Rising velocity of the patches $V^+$ as a function of time $t$. The Bond number is ranging from 15 to 35, which corresponds to bubbles with a spherical cap radius between 9 mm and 13 mm. Error bars correspond to the standard deviation measured by averaging over 10 patches. (b) Rising velocity of the patches $V^+$ as a function of the film thickness $h$. The error for $h$ is taken to be 50 nm (the choice of this value has no noticeable impact on the errors calculated for $\eta_s$ and $V_d$, see next figures).} 
 \label{fig:Vplus=f(t)}
\end{figure}

\begin{figure}[htb]
 \centering
   \includegraphics{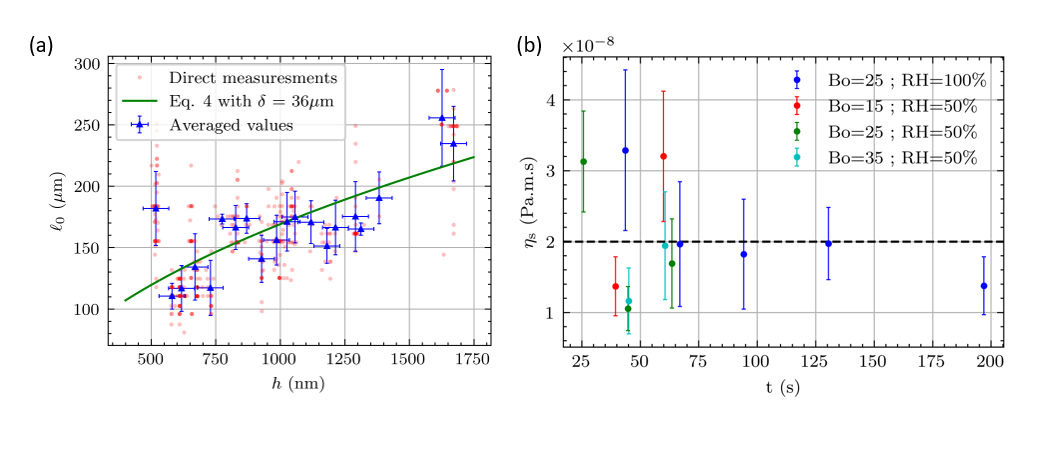}
 \caption{(a) Evolution of the characteristic size of the patches $\ell_0$ ($\mu$m) as a function of the film thickness $h$ (nm) for 387 measurements performed on all systems. The orange circles correspond to direct measurements. The blue triangles correspond to the average values of $\ell_0$ over an interval of 50 nm for $h$. The green curve represent an adjustment of Eq. \ref{lambda} with $\delta$ = 36 $\mu$m. Vertical errorbars correspond to the standard deviation measured. (b) Surface shear viscosity $\eta_s$ as a function of time $t$. The solid line corresponds to the mean value measured for the surface shear viscosity: $\eta_s = (2.0 \pm 0.1). 10^{-8}$ Pa.m.s. The error bars take into account errors in the measured parameters and are calculated using the error propagation formula. 
 }
 \label{fig:eta_s=f(t)_l0=f(h)}
\end{figure}

As shown in Fig. \ref{fig:Vplus=f(t)}(a), the mean values of $V^{+}$ decrease as a function of the age $t$ of the bubbles since their generation whatever the bubble sizes (Bo between 15 and 35 at 50 $\%$ humidity and Bo = 25 at 100 $\%$ humidity). The rising velocity $V^{+}$ at a given time is furthermore expected to depend essentially on the film thickness $h$, itself very dependant on the relative humidity RH after a few tens of seconds. This is illustrated on figure \ref{fig:Vplus=f(t)}(b), which shows that plotting $V^{+}$ versus $h$ allows to reasonably rescale the data. The corresponding sizes $\ell_0$ of the patches are reported in Fig. \ref{fig:eta_s=f(t)_l0=f(h)}(a) as a function of the film thickness $h$. Their values show a wide dispersion. Nevertheless, the averaged values tend to grow with the film thickness. 

The obtained values of $V^+$ can be compared to the model proposed by Seiwert \textit{et al.} \cite{Seiwert2017} for planar films. Here, this model, based on a force balance and on surface conservation, is adapted to the geometry of surface bubbles. 

First, the authors propose that the driving force is gravity, since the lower thickness of the patches acts as 2D buoyancy \cite{adami2014capillary}. 
This driving force is balanced by the viscous dissipation due to the surface shear viscosity $\eta_{\text{s}}$. 
The shear is due to the difference between $V^+$ and $V^-$, the respective velocities of the rising and descending patches. For surface bubbles (Fig.\ref{fig:Schéma_dispositif_experimental-Crop_bulle_monochrome} (a)), this force balance writes: 
\begin{equation}
2 \eta_{\text{s}}  \frac{(V^{+}-V^{-})}{\ell_{\text{0}}^{2}} \ \sim \ \rho g\sin{\alpha}(h-h_{\text{p}}),  \label{eq:ForceBalance}
\end{equation} 
where $h_{\text{p}}$ is the thickness of the patches. 
The factor 2 accounts for the two sheared interfaces on both sides of the film. 

Second, $V^-$ can be related to $V^+$ 
assuming surface conservation: the surface of rising thin patches exactly compensates the surface of the thick descending patches.
In the bubble geometry, the total rising surface $dS^+$ crossing a reference horizontal line at the top of the meniscus during a time interval $dt$ is
$dS^+ = V^+ dt \ \times \ \pi n_0  \ell_0^2 \ \times \ 2\pi r_0$
where $r_0$ is the radius of the bubble on this line. This rising surface must be exactly compensated by the surface of descending liquid 
$dS^- = V^- dt \ \times \ (1 - \ \pi n_0  \ell_0^2) \ \times \ 2\pi r_0$, which leads to a prediction for the drainage velocity of the liquid : 
\begin{equation}
    V^- = V^+ \frac{\pi n_0  \ell_0^2}{1 - \pi n_0  \ell_0^2}
    \label{eq:drainage_velocity_V-}
\end{equation}

Combining the force balance from Eq. \ref{eq:ForceBalance} with Eq. \ref{eq:drainage_velocity_V-} and assuming $(h-h_p) \simeq 0.2 h$ \cite{Nierstrasz1998}, we get :
\begin{equation}
  \eta_s  \sim \frac{\rho g h \ell_0^{2}\sin{\alpha}}{10 \ V^+ \left( \frac{1 \ - \ 2 \pi n_0  \ell_0 ^{2}} {1 \ - \ \pi n_0  \ell_0^{2}}\right)}
    \label{eq:SurfaceViscosity}
\end{equation}

The values of $\eta_{\text{s}}$ deduced from  Eq. \ref{eq:SurfaceViscosity} and from our measurements of $h$, $V^{+}$, $\alpha$, $n_{\text{0}}$ and $\ell_{\text{0}}$ are represented in Fig. \ref{fig:eta_s=f(t)_l0=f(h)}(b). 
They  feature a constant behaviour with a surface shear viscosity around $\eta_s \simeq 2.0 \cdot 10^{-8}$ Pa.m.s whatever the size of the bubbles or the relative humidity, which confirms the reliability of this approach.
The values of surface shear viscosities in surfactant-based systems are controversial. Zell \textit{et al.} cannot measure it and conclude that it must be smaller than $10^{-8}$ Pa.m.s. Other, mostly indirect measurements propose values in the range of 10$^{-6}$ - 10$^{-8}$ Pa.m.s \cite{Stevenson2005,pitois2005liquid}. Our values thus point to the lower limit of the indirect measurements and is slightly higher than the maximum value inferred by Zell \textit{et al.}
Note that here the bulk viscosity is neglected compared to the surface shear viscosity ($\eta h$ $\sim 10^{-9}$~Pa.m.s, with $\eta$ the bulk viscosity that is approximately that of water and $h$ the film thickness).

Let us now focus on the destabilization of the pinch, by concentrating on the patches size.
To predict $\ell_0$, Shabalina \textit{et al.} \cite{Shabalina2019} proposed a Rayleigh-Taylor like description \cite{Chandrasekhar}, which we will compare to our measurements. 
In this model, the gravity difference between the pinch \cite{Aradian2001} and the overlaying thick, heavier film, is the motor of the instability. 
At the frontier between the thin and the thick film, the thickness transition has an extension $\delta$ (see \cite{Shabalina2019} for more details), which leads to an extra surface and a subsequent line tension \cite{Couder1989}.
Among the stabilizing mechanisms proposed, we assume that this line tension is the main stabilizing parameter.
Making use of these assumptions leads to the following prediction for the wavelength $\lambda$ of the instability :

\begin{equation}
    \lambda (h) \sim \ell_{\text{0}}(h) = \left( \frac{\gamma (h-h_p)}{\rho g \delta} \right)^{1/2}
    \label{lambda}
\end{equation}

This prediction is represented against our data in Fig. \ref{fig:eta_s=f(t)_l0=f(h)}(a) and features a good agreement provided $\delta \sim $ 36 $\mu$m, \textit{ie} of the order of the patches size. 
This order of magnitude is in agreement with the observation that a few interference fringes can be seen inside a patch. This is slightly visible in Figure \ref{fig:Photo_bulle_couleurs} but much more convincing in the Figure 6 of Lhuissier \textit{et al.} \cite{Lhuissier2012}. In their thicker films, Shabalina \textit{et al.} \cite{Shabalina2019} also measure a transition length $\delta$ around 100 $\mu$m. This length scale is also in agreement with that measured in another article by Seiwert \textit{et al.}\cite{seiwert2013extension}.
Finally, the underlying mechanism, which fixes the transition length $\delta$ is still unknown but our measurement gives an order of magnitude in line with other observations in the literature.

Note that this representation assumes that the film thickness within the bubble cap, away from the pinched zone, features negligible variations. 
This is coherent with the constant hole opening velocity reported in different works \cite{PoulainPRL2018,Champougny2016} - even on metric bubbles \cite{Cohen2017} - that make use of the roughly homogeneous thickness to retrieve the film thickness from the so-called Taylor-Culick velocity \cite{culick1960}.

In addition, the dependency $\ell_{\text{0}} \propto h^{1/2}$, also found upon matching the curvature of the spherical cap to that of the pinch by Howell  \cite{howell1999draining} and Lhuissier \textit{et al.} \cite{Lhuissier2012} is recovered.
This supports the hypothesis of a gravity-driven instability in good agreement with the qualitative observation that the marginal regeneration cannot be observed anymore with a bubble pending under a wet solid plate.

\section{Drainage due to marginal regeneration}

\begin{figure}[h!]
 \centering
   \includegraphics[width=.6\linewidth]{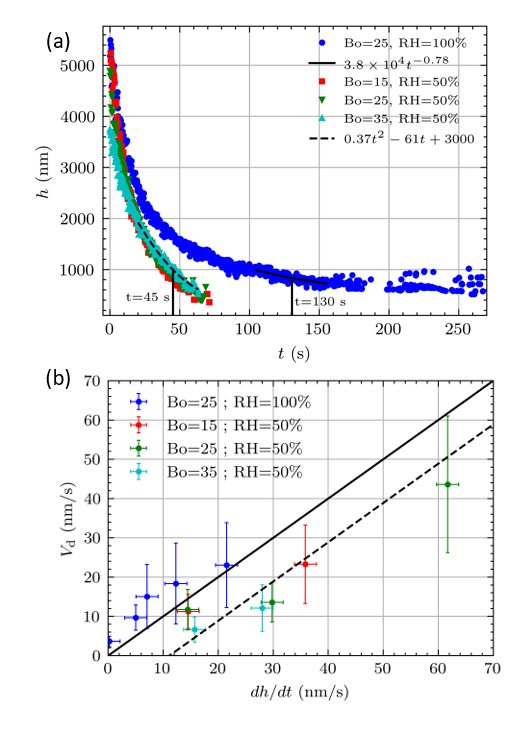}
 \caption{(a) Time evolution of the film thickness $h$ at the bubble apex for different bubble sizes and humidities. The dashed and dotted black lines correspond respectively to second-order and power law adjustment of the film thickness $h$ as a function of time $t$ around the instant $t$ = 45 s for Bo = 35 obtained at RH=50 $\%$ and $t$ = 130 s for bubbles with Bo = 25 obtained at RH=100 $\%$. (b) Thinning rate $V_d$ predicted by the model as a function of the measured thinning rate $dh/dt$. The black line corresponds to a line of equation $y$ = $x$. The black dashed line corresponds to the equation $y$ = -11.1 + $x$. The error bars take into account errors in the measured parameters and are calculated using the error propagation formula.}
 \label{fig:h=f(t)_Vd=f(dhdt)}
\end{figure}

Finally, we now compare the drainage induced by the sliding puzzle dynamics to direct measurements of the film thinning rate measured by spectrometry, as mentioned in the preceding section. 
The time evolution of $h$ is plotted in Fig.\ref{fig:h=f(t)_Vd=f(dhdt)}(a).

As expected, the variation of the thickness $h$ along time is in good agreement with the model presented in previous works (not shown, see references \cite{PoulainPRL2018,Miguet2020,Lhuissier2012}). The data plotted obtained at an atmospheric humidity of 50 \% are actually extracted from Miguet \textit{et al.} \cite{Miguet2020}.

From these curves, we can deduce the thinning rate of the film at the top of the bubble cap  at different times by locally making a power law or a second-order adjustment respectively in saturated environment or in presence of evaporation. Their derivative gives the value of $dh/dt$ (see the fitting in Fig.\ref{fig:h=f(t)_Vd=f(dhdt)}(a) at time $t$ = 45 or 130 s for example).
The model of Lhuissier et al. \cite{Lhuissier2012} and its extensions \cite{PoulainJFM2018,Miguet2020} provide a good scaling for these data but the most appropriate quantification of $h$ and $dh/dt$ are given by these local fits. The corresponding equation given in the caption of figure \ref{fig:h=f(t)_Vd=f(dhdt)}(a) are indicative. Note that the reason why the range of thinning rates is limited for the 100 \% humidity experiment is that the thinning is slower because of the lack of evaporation. In principle, it could be possible to reach faster thinning rates at very short time. Nevertheless, the patches are not well developed at the beginning of the experiment and it is impossible to extract reliable measurements.
Finally, the thinning rate $V_d$ due to the marginal regeneration is linked to $V^-$ by the mass conservation so that $V_d \sim \frac{V^- \ h}{R}$.

In Fig. \ref{fig:h=f(t)_Vd=f(dhdt)}(b), we compare both thinning rates. 
In the absence of evaporation (at RH=100 $\%$), the measured thinning rates are in excellent agreement. 
This means that the sliding puzzle dynamics entirely describes the drainage of a bubble over time. 
Marginal regeneration is therefore the main thinning mechanism in a saturated humidity environment.

For RH=50 \%, the thinning rates measured directly at the top of the bubble cap are more important than the ones attributed to marginal regeneration, which is due to the contribution of evaporation to the thinning of the spherical cap. 
Adding the evaporation flux to the thinning rate predicted by the marginal regeneration model should allow to recover the actual thinning rate. From our measurements in Fig. \ref{fig:h=f(t)_Vd=f(dhdt)}(b), we deduce an evaporation flux of 11 nm.s$^{-1}$.
This is smaller but close to the flux calculated taking into account the convective evaporation of the liquid bath \cite{Miguet2020,Dollet2017}, which is around 35 nm.s$^{-1}$.

Note that the main approximation is the use of circles to estimate the area of the patches. The true, elongated, shape of the patches would result in a decreased density of presence of thin zones along the virtual line where the surface conservation is used to estimate the downward velocity $V^{-}$ of the thick zones. This would slightly lower the downward velocity (equation \ref{eq:drainage_velocity_V-}) and the corresponding thinning rate $V_d$. The second important approximation is that we compare the thinning rate corresponding to the \textit{sliding puzzle dynamics} to that measured at the apex of the bubble. As mentioned earlier, the underlying assumption that this thinning rate at the apex is representative of that at the foot of the bubble is corroborated by previous works on the drainage of surface bubbles \cite{Lhuissier2012} as well as by the constant opening velocity of the hole after puncture.

\section{Conclusion}
In conclusion, our data allow for the first time to quantify directly the effect of marginal regeneration on the overall drainage of a bubble cap. Our results agree with a drainage model where marginal regeneration explains completely the thinning dynamics of the whole cap in the absence of evaporation.

To some extent, this work can be seen as a confirmation of the assumption of Lhuissier \textit{et al.}\cite{Lhuissier2012} that the drainage through the pinch and the marginal regeneration are coupled and of the same order of magnitude. This is also a counterpart of this model in the sense that our results suggest that both contributions are not of the same order of magnitudes but actually equal because the possibility for thick zones to descend is conditioned by the ascension of thin patches, in agreement with the view of Seiwert \textit{et al.} \cite{Seiwert2017} for foam films.
The scenario that we propose is thus the following.
The liquid drainage is indeed limited by the pinch thickness.
Nevertheless, this liquid flow necessitates a replacement of the thick zones flowing downwards by thinner zones rising upwards to ensure surface conservation.
A degree of freedom to ensure the global coherence of the problem is the thickness of the pinch, which impacts both the sliding puzzle dynamics and the limitation of the liquid flow.
Capillary suction therefore leads to the pinching of the film, which limits the liquid velocity, while marginal regeneration regulates the thickness of this pinching and allows the sliding puzzle dynamics.
Additionally, we showed that the rising dynamics of the patches are well described by the model of Seiwert \textit{et al.} \cite{Seiwert2017} driven by gravity and limited by surface shear viscosity. The liquid film forming the cap of the bubble can thus be viewed as two-dimensional fluid with a local density proportional to its thickness. The motions induced by the density differences within the film are limited by surface shear viscous dissipation. The overall thinning of this 2D sheet is conditioned by the 3D Poiseuille flow within the pinch since it induces the thinning of the pinch and eventual detachment of thin patches.
Lastly, our data are in good agreement with a gravity-driven instability \cite{Shabalina2019}. 
We believe that the experimental description proposed in this letter is a necessary step to describe further the bubble drainage, for example for other surfactants or at larger concentrations.

\section*{Acknowledgments}
We are grateful to Fran\c{c}ois Boulogne for fruitful discussions about evaporation and for the design of the oospectro library, to I. Cantat and E. Shabalina for constructive discussions and to F. Restagno and A. Salonen for careful reading and useful advice. 
We thank S. Guichard for the professional photographs of our bubbles. 
Funding from ESA (MAP Soft Matter Dynamics)
and CNES (through the GDR MFA) is acknowledged.
Author Y. Fang is employed by PepsiCo R\&D. 
This study was funded by PepsiCo R\&D. 
The views expressed in this article are those of the authors and do not necessarily reflect the position or policy of PepsiCo, Inc.

\bibliographystyle{unsrt}

\end{document}